\documentclass[lettersize,journal]{IEEEtran}
\usepackage{amsmath,amsfonts}
\usepackage{algorithm}
\usepackage{array}
\usepackage[caption=false,font=normalsize,labelfont=sf,textfont=sf]{subfig}
\usepackage{textcomp}
\usepackage{stfloats}
\usepackage{url}
\usepackage{verbatim}
\usepackage{graphicx}
\usepackage{cite}
\usepackage{booktabs} 
\usepackage{threeparttable} 
\usepackage{pifont} 

\IEEEoverridecommandlockouts
\usepackage{algorithmicx}
\usepackage{pdfcomment}
\usepackage{bookmark}
\usepackage{enumerate}
\usepackage{bm}
\usepackage{amsmath,epsfig,amssymb,amsfonts,color}
\newcommand{\reffig}[1]{Fig. \ref{#1}}

\usepackage{multirow}
\newcommand{\tabincell}[2]{\begin{tabular}{@{}#1@{}}#2\end{tabular}}  

\begin{document}

\title{Measurement-Based Small-Scale Channel Model for Sub-6 GHz RIS-Assisted Communications}

\author{Jian Sang,~\IEEEmembership{Graduate Student Member,~IEEE,} Jifeng Lan,~\IEEEmembership{Graduate Student Member,~IEEE,} Mingyong Zhou,~\IEEEmembership{Graduate Student Member,~IEEE,} Boning Gao, Wankai Tang,~\IEEEmembership{Member,~IEEE,} Xiao Li,~\IEEEmembership{Member,~IEEE,} Michail Matthaiou,~\IEEEmembership{Fellow,~IEEE,}  Shi Jin,~\IEEEmembership{Fellow,~IEEE,} and Marco Di Renzo,~\IEEEmembership{Fellow,~IEEE} 

\thanks{Copyright (c) 2015 IEEE. Personal use of this material is permitted. However, permission to use this material for any other purposes must be obtained from the IEEE by sending a request to pubs-permissions@ieee.org.}
\thanks{Manuscript received August 22, 2023; revised January 04, 2024, accepted February 02, 2024. This work was supported in part by the National Natural Science Foundation of China under Grants 62231009 and 62261160576; in part by the Natural Science Foundation of Jiangsu Province under Grant BK20211511; in part by the Start-up Research Fund of Southeast University under Grant RF1028623267; and in part by the joint project of China Mobile Research Institute and Southeast University. The work of M. Matthaiou was supported by a research grant from the European Research Council (ERC) under the European Union's Horizon 2020 research and innovation programme (grant No. 101001331). The work of M. Di Renzo was supported in part by the Horizon Europe projects COVER-101086228, UNITE-101129618, and INSTINCT-101139161, and the ANR projects NF-PERSEUS 22-PEFT-004 and PASSIONATE ANR-23-CHR4-0003-01. \textit{(Corresponding author: Xiao Li.)}}
\thanks{Jian Sang, Jifeng Lan, Mingyong Zhou, Boning Gao, Wankai Tang, Xiao Li, and Shi Jin are with the National Mobile Communications Research Laboratory, Southeast University, Nanjing 210012, China (email: \{sangjian, lanjifeng, myzhou, gaobn, tangwk, li\_xiao, jinshi\}@seu.edu.cn).}

\thanks{Michail Matthaiou is with the Centre for Wireless Innovation (CWI), Queen’s University Belfast, Belfast BT3 9DT, U.K (email: m.matthaiou@qub.ac.uk).}
\thanks{Marco Di Renzo is with the Université Paris-Saclay, CNRS, CentraleSupélec, Laboratoire des Signaux et Systèmes, Gif-sur-Yvette 91192, France (email: marco.di-renzo@universite-paris-saclay.fr).}
}

\maketitle

\begin{abstract}
Reconfigurable intelligent surfaces (RISs) have attracted increasing interest from both academia and industry, thanks to their unique features on controlling electromagnetic (EM) waves. Although theoretical models for RIS-empowered communications have covered a variety of applications, yet, very few papers have investigated the modeling of real propagation characteristics. In this paper, we fill this gap by providing an empirical statistical channel model to describe the small-scale channel variations for an RIS-assisted broadband system at 2.6 GHz. Based on real channel measurements in outdoor, indoor and outdoor-to-indoor (O2I) environments, we compare and analyze the global, inter-cluster and intra-cluster parameters. Measurement results indicate that the deployment of an RIS with proper phase configurations can significantly improve the channel quality by enhancing the $K$-factor and reducing the time dispersion. The small-scale fading is well characterized by the proposed statistical model and the empirical channel parameters. These results are essential for the design of emerging RIS-assisted wireless systems for future applications.
\end{abstract}


\IEEEpeerreviewmaketitle

\section{Introduction}
Reconfigurable intelligent surface (RIS) has recently been considered as one of the promising solutions for underpinning the future of wireless communications \cite{Matthaiou, bx1, Degli}. Generally, an RIS refers to a nearly-passive two-dimensional metasurface, where several programmable unit cells are embedded periodically \cite{Zhang,b1,br1}. By independently acting as diffuse scatterers but jointly performing beamforming in a desired direction, these unit cells can be regarded as an energy-efficient alternative to large-scale multiple-input multiple-output (MIMO) systems. Thanks to their ability of controlling the propagation of electromagnetic (EM) waves in an adaptive manner, RISs can improve the channel quality or even shape a communication-friendly channel \cite{b2, feng1, feng2}.

Up to now, there have been some works developing deterministic and statistical channel models. The RIS-aided bipartite cascaded Rician channel was introduced in \cite{ba}. It is constituted by the transmitter (Tx)-RIS and RIS-receiver (Rx) channels having different $K$-factors (KFs). The authors of \cite{ba} derived a tight upper bound on the ergodic spectral efficiency. In \cite{bb}, a three-dimensional (3D) geometry-based stochastic model (GBSM), inspired by the existing 3GPP standardized modeling framework, was generalized for application to RIS-assisted channels. In particular, the traditional Tx-Rx small-scale fading parameters were extrapolated for application to an RIS cascaded channel. Based on a 3D cylindrical model, the authors of \cite{bc} proposed a theoretical 3D RIS-assisted MIMO channel model, which considered line-of-sight (LOS) as well as the single-bounced and double-bounced modes. Surprisingly, although the theoretical modeling of RIS-aided channels has been investigated in the literature, the empirical characterization of small-scale fading based on real measurements has not been reported.

In this study, we investigate the small-scale statistical channel characteristics of an RIS-assisted wireless system based on a channel measurement campaign performed at 2.6 GHz. Three typical communication scenarios, including outdoor, indoor and outdoor-to-indoor (O2I) environments, are considered. In each scenario, three propagation modes, including intelligent reflection with RIS (IRWR), specular reflection with RIS (SRWR), and without RIS (WR), are investigated. We formulate a statistical channel model for describing the small-scale propagation behavior of EM waves. Then, based on the measured channel data, we characterize the global, inter-cluster, and intra-cluster statistics. It is found that the KF of the IRWR mode is greatly higher, compared to the SRWR and WR modes. Moreover, the root mean square delay spread (RMS DS) of a cluster is evaluated, unveiling that the IRWR mode can effectively focus the signal energy and reduce the time dispersion.

\begin{figure}[ht]
\centering
\includegraphics[width=3in]{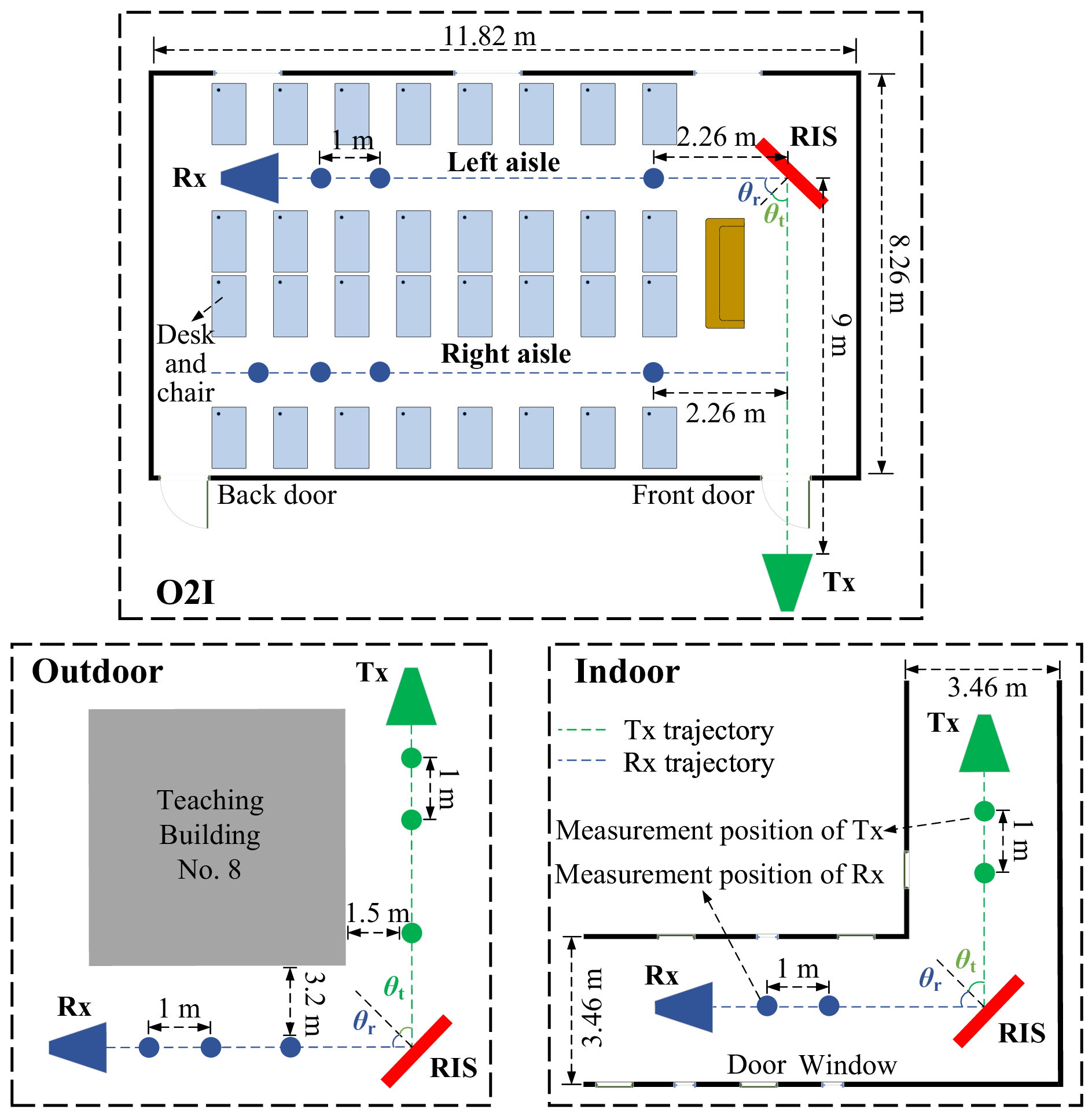}
\caption{Schematic diagram of the channel measurements.}
\label{fig1}
\vspace{-0.3cm}
\end{figure}

\section{Measurements of RIS-assisted channels}

In this paper, we focus on an RIS-assisted single-input single-output (SISO) system.  The direct link between the Tx and Rx is in non-line-of-sight (NLOS), and the RIS is deployed to establish a virtual cascaded LOS link. However, in real environments, a number of multi-path components (MPCs) are also created due to the presence of reflections or scattering, which need to be characterized. In this section, we describe a channel measurement campaign, which was conducted in various scenarios, to investigate RIS-assisted NLOS communications.

\subsection{Channel measurement campaign}
The measurement system is comprised of a vector network analyzer (VNA), a power amplifier (PA), a low noise amplifier (LNA), radio frequency (RF) cables, and two directional horn antennas, as well as a fabricated RIS with central frequency of 2.6 GHz. The RIS is comprised of 32 unit cells per column and 16 unit cells per row, with a physical size of 1.6 m × 0.8 m. Each unit cell can be independently programmed with a 1-bit phase resolution, i.e., it has two coding states identified by the coding ``0'' and coding ``1'', which correspond to the PIN diode being in the ``off'' state and ``on'' state, respectively. A 190 MHz bandwidth signal ranging from 2.5 GHz to 2.69 GHz, with 191 scanning points included, is selected as the broadband signal. The port 1 of the VNA is used to transmit the signal and the port 2 to receive the signal. The scattering parameter $S_{21}$ is collected as the channel transfer function (CTF) in the frequency domain. 

As shown in \reffig{fig1}, the channel measurements for the RIS-assisted SISO system are conducted in three scenarios, including the outdoor, indoor, and O2I environments, in the Jiulonghu Campus of Southeast University, in Nanjing, China. The outdoor measurement is conducted at the building corner next to the Teaching Building No. 8. In this scenario, the Tx and Rx are placed at the east and south sides of the building respectively, so that the communication links is in NLOS, due to the blockage by the exterior wall of the building. The RIS is deployed at the corner of the building to improve the coverage. The indoor measurement is carried out in the corridor on the first floor of the Teaching Building No. 2, where both sides of the corridor are perpendicular to each other, i.e., the link is in NLOS. In this scenario, the Tx and Rx are placed in the south-north corridor and the east-west corridor respectively, and the RIS is deployed at the intersection of these two corridors. The O2I measurement is conducted in the classroom on the first floor of the Teaching Building No. 7, where the Tx is fixed outside the classroom and faces the front door, which is kept open during the measurements. The Rx is moved along the left aisle and right aisle in the classroom, and the RIS is deployed at the intersection of the Tx-front door line and the left aisle.

Let $\theta_t$ and $\theta_r$ denote the angle of arrival from the Tx to RIS and the angle of departure from the RIS to Rx, respectively. In particular, for the channel measurements in the outdoor and indoor scenarios, we have $\theta_t = \theta_r = 45^\circ$, which means that the Tx and Rx are located at the mirror positions with respect to the RIS. The start and end measurement distances from the Tx to RIS as well as the start and end measurement distances from the RIS to Rx are $5$ m, $18$ m, $5$ m, $18$ m respectively, with a step of $1$ m. Therefore, $14 \times 14 = 196$ points are measured in each of these two scenarios. In the O2I scenario, the positions of the Tx and RIS are fixed at a distance of $9$ m between them and $\theta_t=45^\circ$. Meanwhile, in the left aisle, we have $\theta_r = \theta_t = 45^\circ$, and the start and end distances from the RIS to Rx are $2.26$ m and $10.26$ m respectively. In the right aisle, we have $\theta_r \ne \theta_t$, and the start and end distances are $2.26$ m and $10.26$ m, which denote the perpendicular distance from the Rx to the Tx-RIS line. In both the left and right aisles, the movement step of the Rx is $1$ m. As a consequence, $ 9 + 9 =18 $ points are measured in the O2I scenario.

In each scenario, we consider three propagation modes: IRWR, SRWR, and WR modes. In the IRWR mode, the RIS performs phase optimization according to the ``Dynamic Threshold Phase Quantization (DTPQ)'' coding scheme \cite{bd}, to focus the signal energy towards the intended location. In the SRWR mode, the RIS does not perform phase optimization and all the unit cells of the RIS are uniformly configured to the coding state ``0'', which can be viewed as the equivalent of an equal-sized metal plate \cite{tang_twc, be}. In the WR mode, the RIS is removed, and the other measurement configurations remain unchanged, and, hence, only the inherent NLOS link between the Tx and Rx exists. A detailed description of the channel measurement campaign can be found in \cite{bf}. 

\subsection{Measurement data post-processing}
The data $S_{21}$ collected by the VNA is denoted by $H_V(f)$, and it accounts for the channel response $H(f)$, the Tx antenna response $G_t(f)$, the Rx antenna response $G_r(f)$, and the system response $G(f)$. Thus, through a back-to-back calibration, the channel response $H(f)$ is given by
\begin{equation}
\label{eq1}
H(f) = {H_V}(f)/(G(f)G_t(f)G_r(f)).
\end{equation}
Then, the channel impulse response (CIR) in the time domain can be calculated by the inverse Fourier transform, as follows:
\begin{equation}
\label{eq2}
h(t,\tau ) = {\rm{IFT}}(H(f) \times {{\rm{W}}_{{\rm{hann}}}}),
\end{equation}
where $\rm{IFT(\cdot)}$ denotes the inverse Fourier transform, and ${\rm{W}}_{{\rm{hann}}}$ represents the Hanning-window. Then, the power delay profile (PDP) is calculated by
\begin{equation}
\label{eq_PDP}
{PDP} = {\left| {h(t,\tau )} \right|^2}.
\end{equation}
The noise may be included in \eqref{eq2} and \eqref{eq_PDP}, thus, the actual CIR and PDP need to be estimated by detecting valid MPCs and eliminating the noise. The detection method can be found in \cite{bf}, and is omitted here due to space limitations.

\begin{figure*}[t]
\centering
\includegraphics[height=2in]{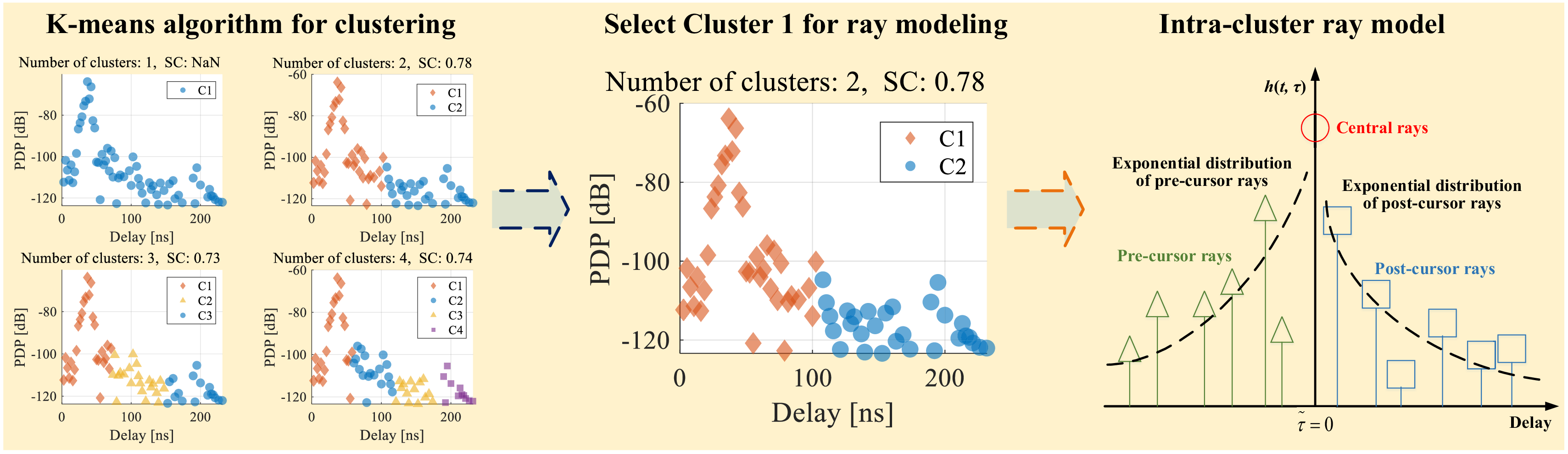}
\caption{Processing flow for identifying clusters and rays.}
\label{fig2}
\end{figure*}

\section{Statistical modeling of RIS-assisted channels}
In this section, we propose a temporal small-scale channel model, encompassing global, cluster-level and ray-level parameters, to characterize the CIR for the considered RIS-assisted wireless channel. Also, the inter-cluster and intra-cluster statistics of the proposed model are characterized.

\subsection{Small-scale channel model}
The small-scale channel characterization has been inherently integrated into previous 3D statistical channel models (see for instance \cite{bg} and references therein). In general, statistical channel models utilize the GBSM, which can capture the fast-fading channel characteristics. Considering the many unit cells on an RIS, its propagation behavior can be viewed as equivalent to that of a massive MIMO system. Thus, based on the WINNER II and Saleh-Valenzuela (SV) models, we represent the CIR of the RIS-assisted channel as
\begin{equation}
\label{eqr1}
\begin{aligned}
h(t,\tau ) & = \sqrt {\frac{{K(t)}}{{K(t) + 1}}} {h_{LOS}}(t)\delta (t - {\tau _{LOS}}(t))
\\ 
& + \sqrt {\frac{1}{{K(t) + 1}}} \sum\limits_{m = 1}^M {\sum\limits_{n = 1}^{{N_m}} {{h_{m,n}}(t)} } \delta (t - {\tau _m}(t) - {\tau _{m,n}}(t)) 
\\
& + \sum\limits_{p = 1}^P {\sum\limits_{q = 1}^{{Q_p}} {{h_{p,q}}(t)} } \delta (t - {\tau _p}(t) - {\tau _{p,q}}(t)).
\end{aligned}
\end{equation}
In \eqref{eqr1}, the RIS-assisted channel is divided into three terms.\footnote{In RIS-assisted systems, more subtle channel classifications may be needed, including the Tx-RIS sub-channel and RIS-Rx sub-channel. However, due to the nearly passive nature of RISs, the parameters of these sub-channels are difficult to be directly observed or measured, requiring more advanced estimation algorithms. This is left to future research work.} The first term is the cascaded virtual LOS component of the Tx-RIS-Rx link, where $K(t)$ is the Rician KF,\footnote{The KF generally refers to the power ratio between the deterministic component and the stochastic components, e.g., the KF can be nonzero in a completely NLOS link. In this paper, the KFs represent the power ratio between the virtual LOS link and the RIS-related MPCs in the IRWR and SRWR modes, and the power ratio between the main reflected path and the other MPCs in the WR mode.} ${h_{LOS}}(t)$ is the amplitude of the cascaded virtual LOS component, while ${\tau _{LOS}}(t)$ is the delay of this cascaded LOS component. The second term denotes the RIS-related MPCs generated from the undesired diffuse scattering in the Tx-RIS link and/or the RIS-Rx link, where $M$ and $N_m$ are the number of clusters and the number of rays in the $m$th cluster, respectively, ${\tau _m}(t)$ is the delay of the $m$th cluster, ${h_{m,n}}(t)$ is the amplitude of the $n$th ray within the $m$th cluster, and ${\tau _{m,n}}(t)$ denotes its delay. The third term indicates the inherent NLOS MPCs of the Tx-Rx link, which are not reflected by the RIS, where $P$ and $Q_p$ are the number of clusters and the number of rays in the $p$th cluster, respectively, ${\tau _p}(t)$ is the delay of the $p$th cluster, ${h_{p,q}}(t)$ is the amplitude of the $q$th ray within the $p$th cluster, with ${\tau _{m,n}}(t)$ denoting its delay.

By recalling that the inherent NLOS MPCs of the Tx-Rx link are usually weak, as Fig. 13 in \cite{bf} showcases, these stochastic MPCs in the third term of \eqref{eqr1} can be neglected to make \eqref{eqr1} easier to understand. Then, for RIS-assisted NLOS communications, a simplified expression of \eqref{eqr1} can be written as given in \eqref{eq3}, which is utilized for the subsequent parameter analysis based on the channel measurement data: 
\begin{equation}
\label{eq3}
\begin{aligned}
h(t,\tau )  & = \sqrt {\frac{{K(t)}}{{K(t) + 1}}} {h_{LOS}}(t)\delta (t - {\tau _{LOS}}(t)) \\ 
& + \sqrt {\frac{1}{{K(t) + 1}}} \sum\limits_{m = 1}^M {\sum\limits_{n = 1}^{{N_m}} {{h_{m,n}}(t)} } \delta (t - {\tau _m}(t) - {\tau _{m,n}}(t)).
\end{aligned}
\end{equation}

\subsection{Inter-cluster parameters}
In this subsection, the K-means algorithm \cite{bh} is utilized for clustering the time-domain MPCs. Traditionally, the K-means algorithm requires to pre-set the number of clusters, whereas this number could change at different measured positions. Therefore, we choose a range of $1 \sim 4$ for the number of clusters, considering its maximum number is less than 4 by visually examining the PDPs. The Silhouette Coefficient (SC) is selected as the evaluation index for the clustering results so as to obtain the optimal number of clusters. Note that $ \rm{SC}\in [-1,1]$, and the larger its value is, the better the clustering performance of the K-means algorithm is. As exemplified in \reffig{fig2}, the SCs of the four clustering modes are $\{NaN, 0.78, 0.73, 0.74\}$, whose maximum value is $0.78$. Thus, in this case, the appropriate number of clusters is assumed to be $2$.

Based on the obtained clustering results, the inter-cluster statistical parameters including the average number of clusters, cluster arrival rate, cluster arrival time, and cluster power decay rate, are investigated. The arrival cluster is assumed to be a Poisson process, where the cluster arrival time, defined as the relative delay difference between two adjacent cluster centers, is modeled by an exponential distribution. The cluster arrival rate is approximated by the inverse of the mean cluster arrival time. The MPC with the highest amplitude within each cluster is selected as its cluster center. The cluster power decay rate is defined as an exponentially decaying function of the amplitudes of the cluster centers.

\subsection{Intra-cluster parameters}
In this subsection, the individual rays in each cluster are identified. The intra-cluster statistical characteristics, including the average number of rays, ray arrival rate, ray power decay time, and intra-cluster RMS DS, are also evaluated.

The intra-cluster ray model proposed in \cite{bi} is used. We consider that the $m$th cluster in the time domain is constituted of a central ray ${\alpha^{(m,0)}}$, a pre-cursor ray set $\left\{ {{\alpha^{(m, - {N_{pre}})}},...,{\alpha^{(m, - 1)}}} \right\}$, and a post-cursor ray set $\left\{ {{\alpha^{(m,1)}},...,{\alpha^{(m,{N_{post}})}}} \right\}$, as demonstrated in \reffig{fig2}. In detail, the ray with the highest amplitude in the cluster is selected as the central ray ${\alpha^{(m,0)}}$. We model the pre-cursor and post-cursor rays as two Poisson processes with arrival rates $\lambda_{pre}$ and $\lambda_{post}$, respectively. The amplitudes $A_{pre}$ and $A_{post}$ of the pre-cursor and post-cursor rays are modeled as exponential functions, with $\gamma_{pre}$ and $\gamma_{post}$ denoting their power decay time respectively, as follows:
\begin{equation}
\label{eq4}
{A_{pre}}(\tau ) = {A_{pre}}(0){e^{\widetilde \tau /{\gamma _{{\rm{pre}}}}}},
\end{equation}
\begin{equation}
\label{eq5}
{A_{post}}(\tau ) = {A_{post}}(0){e^{ - \widetilde \tau /{\gamma _{{\rm{post}}}}}},
\end{equation}
where $\widetilde \tau$ denotes the relative delay of the rays with respect to the cluster center, with $\widetilde \tau = 0$ at the central ray ${\alpha^{(m,0)}}$.

To describe the time-domain shape of the arriving cluster, the intra-cluster RMS DS for the $m$th cluster is defined as
\begin{equation}
\label{eq6}
{\tau _{RMS}} = \sqrt {\frac{{\sum\nolimits_{n = 1}^{{N_m}} {{{\left| {{h_{m,n}}} \right|}^2}{\tau _{m,n}}^2} }}{{\sum\nolimits_{n = 1}^{{N_m}} {{{\left| {{h_{m,n}}} \right|}^2}} }} - {{\left( {\frac{{\sum\nolimits_{n = 1}^{{N_m}} {{{\left| {{h_{m,n}}} \right|}^2}{\tau _{m,n}}} }}{{\sum\nolimits_{n = 1}^{{N_m}} {{{\left| {{h_{m,n}}} \right|}^2}} }}} \right)}^2}}.
\end{equation}

\begin{figure}[t]
\vspace{-0.2cm}
\centering
\includegraphics[width=3.4in]{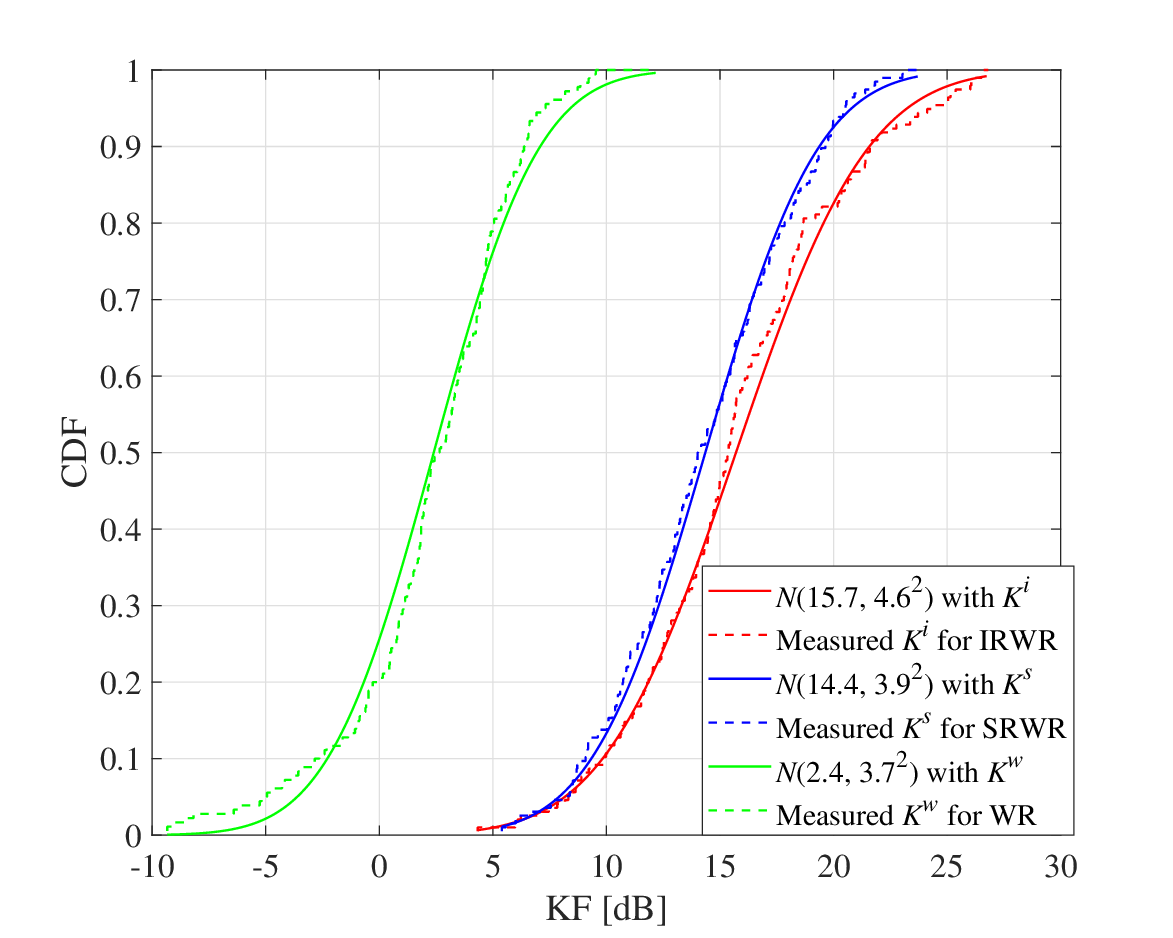}
\caption{KFs in the outdoor scenario.}
\label{fig3}
\vspace{-0.2cm}
\end{figure}

\begin{figure}[t]
\vspace{-0.2cm}
\centering
\includegraphics[width=3.4in]{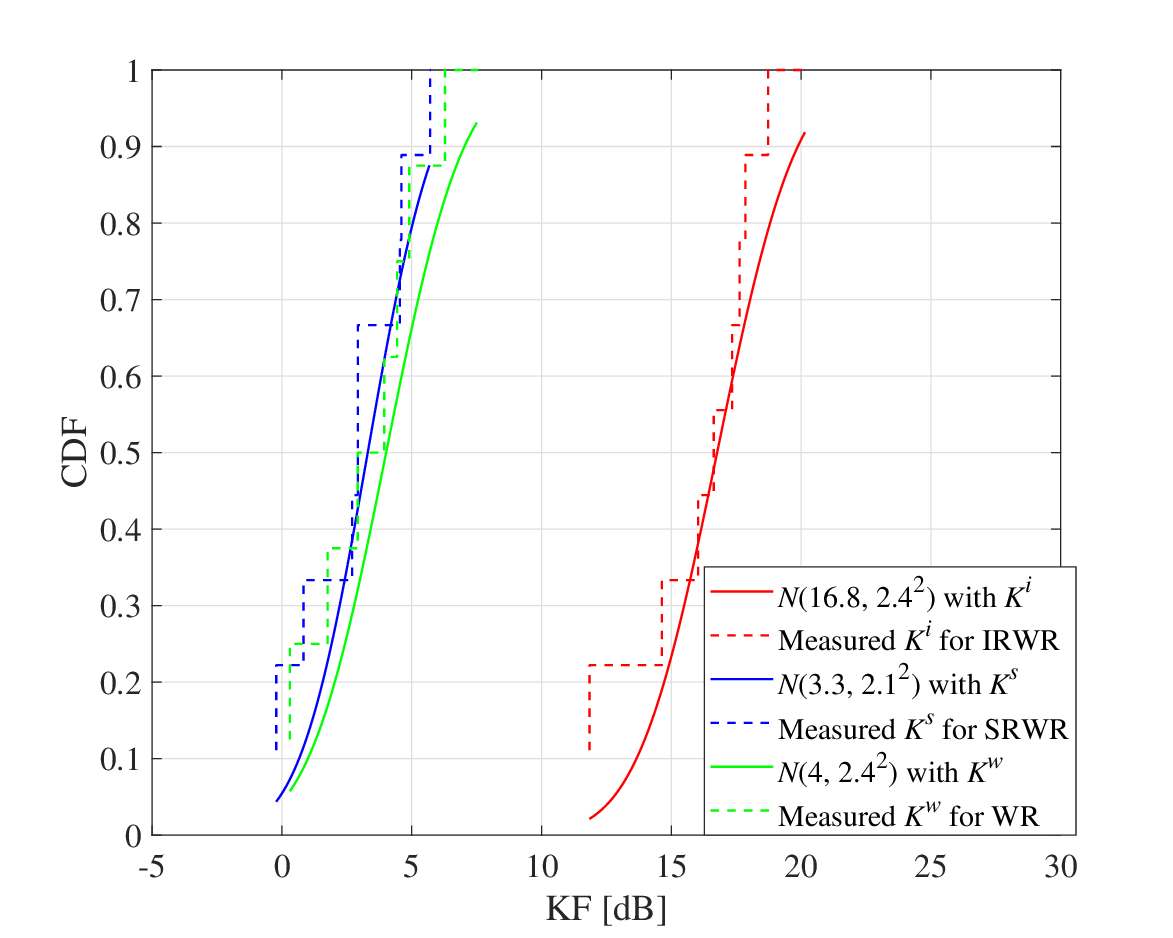}
\caption{KFs in the right aisle in the O2I scenario.}
\label{fig4}
\vspace{-0.2cm}
\end{figure}

\begin{figure}[t]
\vspace{-0.5cm}
\centering
\includegraphics[width=3.6in]{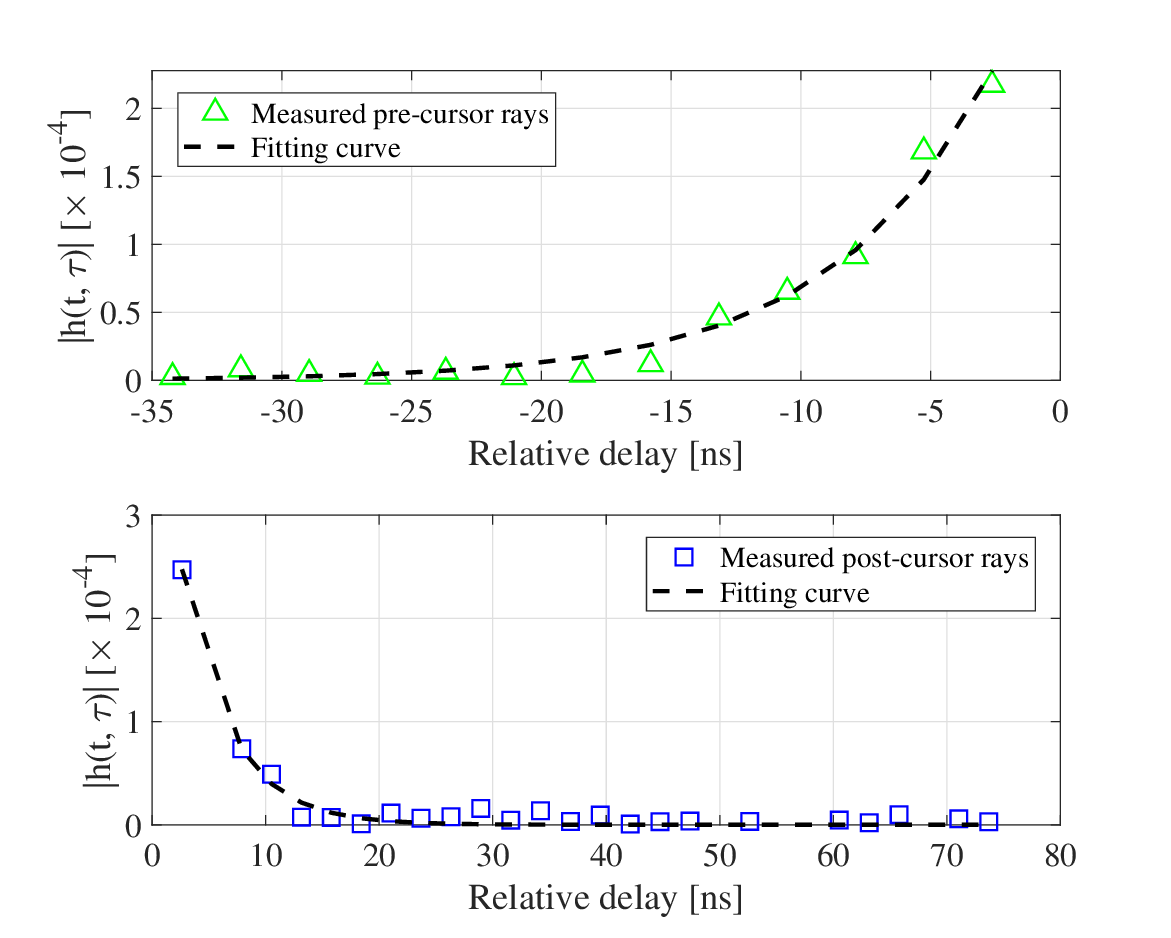}
\caption{Example of the measured rays and the fitting performance.}
\label{fig5}
\vspace{-0.5cm}
\end{figure}

\section{Measurement Results and Analyses}
Based on the measured channel data in the considered three scenarios, the time-domain statistical small-scale behavior is characterized and analyzed in this section. Considering the large measured bandwidth in the sub-6 GHz band, a high delay resolution is ensured for observing in good detail the behavior of the clusters and rays.

\begin{figure*}[ht]
\vspace{-0.5cm}
\centering
\includegraphics[height=2.1in]{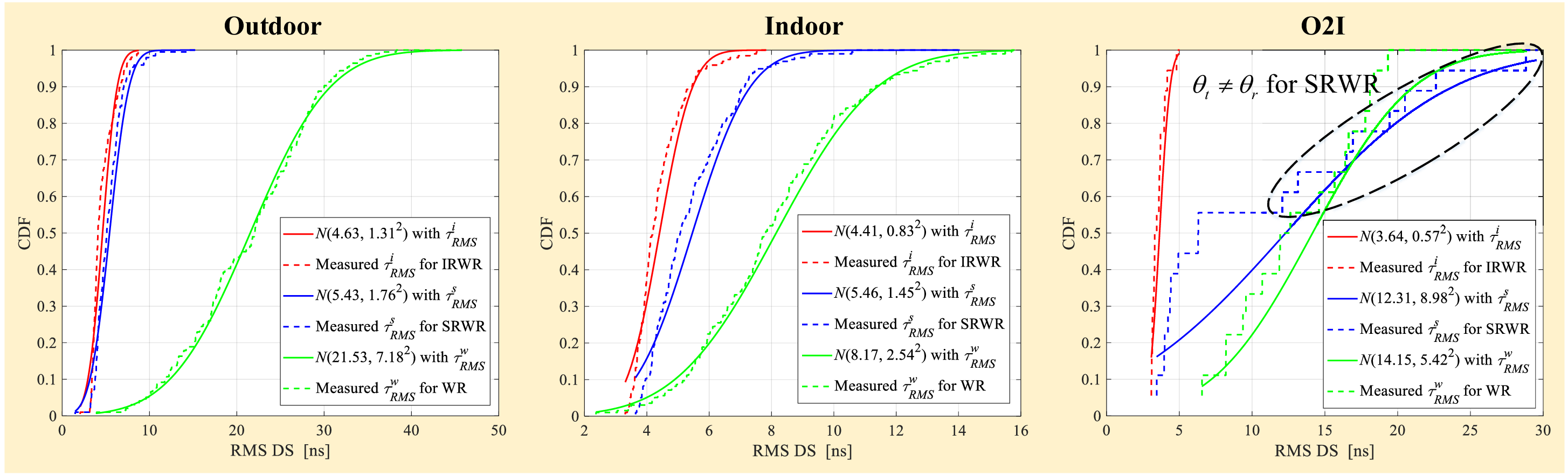}
\caption{Intra-cluster RMS DSs.}
\label{fig6}
\vspace{-0.2cm}
\end{figure*}

\begin{table*}[!hb]
\vspace{-0.2cm}
\centering
\footnotesize
\caption{Summary of the small-scale parameters.}
\label{table_parameter}
\begin{tabular}{|c|c|c|c|c|c|c|}
\hline
\multicolumn{3}{|c|}{\multirow{4}{*}{\tabincell{c}{\textbf{Parameters}}}} 
& \multirow{4}{*}{\tabincell{c}{Outdoor\\ ($\theta_t =\theta_r$)}} & \multirow{4}{*}{\tabincell{c}{Indoor \\ ($\theta_t =\theta_r$)}} 

& \multicolumn{2}{c|}{O2I} \\ \cline{6-7} 
\multicolumn{3}{|c|}{} & & & 
\multirow{3}{*}{\tabincell{c}{Left aisle\\ ($\theta_t =\theta_r$)}}

& \multirow{3}{*}{\tabincell{c}{Right aisle\\ ($\theta_t \ne\theta_r$)}} \\ 

\multicolumn{3}{|c|}{} & & & & \\ 
\multicolumn{3}{|c|}{} & & & & \\ \hline

\multirow{6}{*}{\tabincell{c}{\textbf{Global KF}}} & \multirow{2}{*}{\tabincell{c}{IRWR mode}} & $\mu$ & 15.7 & 12 & 20 & 16.8 \\ \cline{3-7}
& & $\sigma$ & 4.6 & 4.2 & 2.9 & 2.4  \\ \cline{2-7}

& \multirow{2}{*}{\tabincell{c}{SRWR mode}} & $\mu$ & 14.4 & 10 & 13.8 & 3.3 \\ \cline{3-7}
& & $\sigma$ & 3.9 & 4 & 4.4 & 2.1  \\ \cline{2-7}
& \multirow{2}{*}{\tabincell{c}{WR mode}} & $\mu$ & 2.4 & 2 & 1.6 & 4 \\ \cline{3-7}
& & $\sigma$ & 3.7 & 3.5 & 2.2 & 2.4  \\ \hline

\multirow{4}{*}{\tabincell{c}{\textbf{Inter-cluster parameters for IRWR}}} & 
\multicolumn{2}{c|}{Average number of clusters} & 2.3 & 2.2 & \multicolumn{2}{c|}{2.4} \\ \cline{2-7}
& \multicolumn{2}{c|}{Cluster arrival rate [1/ns]} & 0.008 & 0.006 & \multicolumn{2}{c|}{0.012} \\ \cline{2-7}
& \multicolumn{2}{c|}{Cluster arrival time [ns]} & 126.5 & 179.68 & \multicolumn{2}{c|}{85.2} \\ \cline{2-7}
& \multicolumn{2}{c|}{Cluster power decay rate [ns]} & 0.03 & 0.03 & \multicolumn{2}{c|}{0.05} \\ \hline
\multirow{8}{*}{\tabincell{c}{\textbf{Intra-cluster parameters of Cluster 1 for IRWR}}} & 
\multicolumn{2}{c|}{Average number of rays} & 47 & 52 & \multicolumn{2}{c|}{34} \\ \cline{2-7}
& \multicolumn{2}{c|}{RMS DS of cluster [ns]} & 4.63 & 4.41 & \multicolumn{2}{c|}{3.64} \\ \cline{2-7}
& \multirow{3}{*}{\tabincell{c}{Pre-cursor rays}} 
& Power decay time [ns] & 5.62 & 5.56 & \multicolumn{2}{c|}{6.58} \\ \cline{3-7} 
& & Number of rays & 16 & 16 & \multicolumn{2}{c|}{12} \\ \cline{3-7}
& & Arrival rate [1/ns] & 0.27 & 0.29 & \multicolumn{2}{c|}{0.36} \\ \cline{2-7}
& \multirow{3}{*}{\tabincell{c}{Post-cursor rays}}
& Power decay time [ns] & 6.31 & 7.09 & \multicolumn{2}{c|}{6.39} \\ \cline{3-7} 
& & Number of rays & 30 & 35 & \multicolumn{2}{c|}{21} \\ \cline{3-7}
& & Arrival rate [1/ns] & 0.34 & 0.31 & \multicolumn{2}{c|}{0.36} \\ \hline

\end{tabular}
\vspace{-0.2cm}
\end{table*}

\subsection{Global parameters}
The global KFs for the three propagation modes are given in \autoref{table_parameter}, which are fitted to the Gaussian distribution $N(\mu ,{\sigma ^2})$. The moment-based estimation method for the KF in \cite{bj} is used, according to which the wideband channel is divided into different narrowband fading realizations. The measured bandwidth of $190$ MHz is divided into $19$ non-overlapping subbands with each bandwidth of $10$ MHz. From \autoref{table_parameter}, the KF is significantly improved for the IRWR mode, whose mean values are $15.7$ dB, $12$ dB, $20$ dB, $16.8$ dB, compared to $2.4$ dB, $2$ dB, $1.6$ dB, and $4$ dB for the WR mode, in the outdoor, indoor, left aisle of O2I, and right aisle of O2I scenarios, respectively. Also, for the case $\theta_t= \theta_r$, the KF of the SRWR mode is higher than the WR mode and lower than the IRWR mode, yet it falls down to a comparable value to the WR mode when $\theta_t \ne \theta_r$. In the IRWR mode, it is therefore apparent that the signal energy is well focused and the KF is enhanced for an arbitrary angle and distance.

In detail, \reffig{fig3} and \reffig{fig4} show the KFs in the outdoor scenario ($\theta_t= \theta_r$) and in the right aisle of the O2I scenario ($\theta_t \ne \theta_r$) respectively. In these figures, the KFs are well described by Gaussian distributions, where $K^i$, $K^s$, and $K^w$ refer to the KFs for the IRWR, SRWR, and WR modes, respectively. From \reffig{fig3}, the KF of the IRWR mode is higher than that of the SRWR mode, indicating a better signal improvement capability, even though the Tx and Rx are located in the mirror positions with respect to the RIS. Moreover, the KF of the WR mode is the lowest. In \reffig{fig4}, by contrast, the KF of the SRWR mode decreases dramatically, due to its poor beamforming capabilities when the Tx and Rx are located at non-mirror positions with respect to the RIS. These phenomena indicate the critical significance of a proper deployment and phase configuration for the RIS, with the DTPQ method adopted for the IRWR mode serving as a candidate scheme.

\subsection{Inter-cluster parameters}

Considering the better performance provided by the IRWR mode in RIS-assisted channels, its inter-cluster characteristics are mainly analyzed in this subsection. The detailed inter-cluster parameters are summarized in \autoref{table_parameter}. The average number of clusters in the considered three scenarios is about $2$, which indicates a weak MPC effect. This phenomenon may originate from the relatively short measured distance, where the scattering/reflection objects are sparse. This also accounts for their high cluster arrival times of $126.5$ ns, $179.68$ ns, and $85.2$ ns, respectively. The cluster arrival rates are $0.008$ 1/ns, $0.006$ 1/ns, $0.012$ 1/ns and the cluster power decay rates are $0.03$ ns, $0.03$ ns, $0.05$ ns, in the outdoor, indoor, and O2I scenarios, respectively.

\subsection{Intra-cluster parameters}

Based on the clustered PDPs obtained for the IRWR mode, as shown in \reffig{fig2}, and by considering that Cluster 2 (including Clusters 3 and 4, if they exist) is too weak, only the intra-cluster parameters of Cluster 1 are analyzed in this subsection. The measured rays of Cluster 1 and the fitting performance are exemplified visually in \reffig{fig5}, where we can see that both the pre-cursor rays and post-cursor rays are well described by exponential distributions. 

The detailed intra-cluster parameters of Cluster 1 are shown in \autoref{table_parameter}. The intra-cluster RMS DSs are demonstrated in \reffig{fig6}, which are compared across different modes and across different scenarios, where $\tau^i_{RMS}$, $\tau^s_{RMS}$, and $\tau^w_{RMS}$ refer to the intra-cluster RMS DSs for the IRWR, SRWR, and WR modes, respectively. The mean values of the intra-cluster RMS DSs are $\tau^i_{RMS} = 4.63$ ns, $\tau^s_{RMS} = 5.43$ ns, and $\tau^w_{RMS} = 21.53$ ns in the outdoor scenario, $\tau^i_{RMS} = 4.41$ ns, $\tau^s_{RMS} = 5.46$ ns, and $\tau^w_{RMS} = 8.17$ ns in the indoor scenario, as well as $\tau^i_{RMS} = 3.64$ ns, $\tau^s_{RMS} = 12.31$ ns, and $\tau^w_{RMS} = 14.15$ ns in the O2I scenario, respectively. From \reffig{fig6}, it can be seen that the IRWR mode has the lowest mean values across the three scenarios, indicating the weakest time dispersion. The time dispersion for the SRWR mode is stronger than that for the IRWR mode and weaker than that for the WR mode, when $\theta_t=\theta_r$, as indicated for the outdoor and indoor scenarios. By contrast, in the regions where $\theta_t \ne \theta_r$, the performance of the SRWR mode becomes nearly equivalent with that of the WR mode, as shown in \reffig{fig6}.

\section{Conclusion}
In this study, based on empirical channel measurements, we presented a statistical channel model to describe the small-scale behavior of an RIS-assisted broadband system at 2.6 GHz. Three typical communication scenarios, including outdoor, indoor and O2I, were taken into account. In each scenario, three propagation modes, including IRWR, SRWR and WR, were considered to investigate the impact of the deployment of an RIS and its phase configuration on the propagation channel. Multi-level fast-fading parameters including global, inter-cluster and intra-cluster were investigated and compared. Measurement results showcased that the IRWR mode can significantly enhance the $K$-factor and mitigate the time dispersion. The study indicated that the small-scale behavior could be well characterized, by using the parameters estimated from the empirical data. In terms of applications, the results reported in this paper can be used as a reference for the design of RIS-assisted wireless systems and channel models in standardization working groups.





\ifCLASSOPTIONcaptionsoff
  \newpage
\fi

\end{document}